\documentclass[aps,prl,twocolumn,superscriptaddress,footinbib]{revtex4-2}

\usepackage{graphicx}  % Include figure files
\usepackage{subfigure}
\usepackage{dcolumn} % Align table columns on decimal point
\usepackage{bm}% bold math
\usepackage{amssymb}   % for math
\usepackage{amsfonts}   % for math
\usepackage{comment}
\usepackage[colorlinks,linkcolor=blue,anchorcolor=blue,citecolor=blue,urlcolor=blue]{hyperref}
\usepackage{xcolor}
\usepackage{soul}
\usepackage{tabularx,booktabs,ragged2e,multirow}
\newcolumntype{C}{>{\centering\arraybackslash}X}
\usepackage{multirow}
\usepackage{amsmath}
\usepackage{mathrsfs}
\usepackage{mathcomp}
\usepackage{textcomp}
\usepackage{dsfont}
\usepackage{esint}
\usepackage{braket}
\usepackage{textgreek}
\usepackage{lipsum}
\usepackage{marvosym} 
\usepackage{centernot}
\usepackage{cancel}
\usepackage{dashrule}
\usepackage[normalem]{ulem}

\begin{document}

\preprint{APS/123-QED}

%Title of paper
\title{Right-eigenstate-based approach to non-Hermitian superfluidity with two-body loss}

\author{Xuezhu Liu}
\affiliation{Center for Advanced Quantum Studies, School of Physics and Astronomy, Beijing Normal University, Beijing 100875, China}
\affiliation{Key Laboratory of Multiscale Spin Physics (Ministry of Education), Beijing Normal University, Beijing 100875, China}

\author{Ming Lu}
\email{Correspondence: luming@baqis.ac.cn}
\affiliation{Beijing Academy of Quantum Information Sciences, Beijing 100193, China}

\author{Haiwen Liu}
\email{Correspondence: haiwen.liu@bnu.edu.cn}
\affiliation{Center for Advanced Quantum Studies, School of Physics and Astronomy, Beijing Normal University, Beijing 100875, China}
\affiliation{Key Laboratory of Multiscale Spin Physics (Ministry of Education), Beijing Normal University, Beijing 100875, China}
\affiliation{Interdisciplinary Center for Theoretical Physics and Information Sciences, Fudan University, Shanghai 200433, China}

\newcommand{\br}{{\bm r}}
\newcommand{\bk}{{\bm k}}
\newcommand{\bq}{{\bm q}}
\newcommand{\bp}{{\bm p}}
\newcommand{\bv}{{\bm v}}
\newcommand{\bmm}{{\bm m}}
\newcommand{\bA}{{\bm A}}
\newcommand{\bE}{{\bm E}}
\newcommand{\bB}{{\bm B}}
\newcommand{\bH}{{\bm H}}
\newcommand{\bd}{{\bm d}}
\newcommand{\bzero}{{\bm 0}}
\newcommand{\bOmega}{{\bm \Omega}}
\newcommand{\bsigma}{{\bm \sigma}}
\newcommand{\bJ}{{\bm J}}
\newcommand{\bL}{{\bm L}}
\newcommand{\bS}{{\bm S}}
\newcommand\dd{\mathrm{d}}
\newcommand\ii{\mathrm{i}}
\newcommand\ee{\mathrm{e}}
\newcommand\zz{\mathtt{z}}
\makeatletter
\let\newtitle\@title
\let\newauthor\@author
\def\ExtendSymbol#1#2#3#4#5{\ext@arrow 0099{\arrowfill@#1#2#3}{#4}{#5}}
\newcommand\LongEqual[2][]{\ExtendSymbol{=}{=}{=}{#1}{#2}}
\newcommand\LongArrow[2][]{\ExtendSymbol{-}{-}{\rightarrow}{#1}{#2}}
\newcommand{\cev}[1]{\reflectbox{\ensuremath{\vec{\reflectbox{\ensuremath{#1}}}}}}
\newcommand{\red}[1]{\textcolor{red}{#1}} %for displaying red texts
\newcommand{\blue}[1]{\textcolor{blue}{#1}} %for displaying blue texts
\newcommand{\green}[1]{\textcolor{orange}{#1}} %for displaying blue texts
\newcommand{\mytitle}[1]{\textcolor{orange}{\textit{#1}}}
\newcommand{\mycomment}[1]{} %for commenting out
\newcommand{\note}[1]{ \textbf{\color{blue}#1}}
\newcommand{\warn}[1]{ \textbf{\color{red}#1}}

\makeatother

\begin{abstract} We theoretically explore a non-Hermitian superfluid model with complex-valued interaction, inspired by two-body loss stemming from inelastic scattering observed in ultracold atomic experiments. Utilizing both the right-eigenstate-based mean-field theory and its biorthogonal counterpart, we study the properties of the system. Notably, the right-eigenstate-based framework produces smooth and continuous solutions, in stark contrast to the absence of nontrivial solutions and the abrupt discontinuities observed in the biorthogonal-eigenstate-based framework under moderate dissipation. In addition, the lower condensation energy obtained in the former framework suggests its superior suitability for describing this system. 
Furthermore, we explore the impact of backscattering, a crucial factor in realistic systems. Our analysis reveals that, facilitated by two-body loss, even moderate backscattering destabilizes the superfluid state. Sufficiently strong backscattering completely destroys it, highlighting a key mechanism for the fragility of this non-Hermitian quantum phase.

\end{abstract}

\keywords{non-Hermitian superfluid；right-eigenstate-based mean-field theory；two-body loss；backscattering；metastable superfluid}

\maketitle

\mytitle{Introduction}.--Superconducting systems, characterized by their unique macroscopic quantum phenomena and wide-ranging applications in modern quantum technologies, constitute a core area of condensed matter physics \cite{Acín_2018}.
While numerous unconventional and exotic superconductors have been discovered within the framework of closed Hermitian systems \cite{RevModPhys.63.239, RevModPhys.78.17, Stewart1984, RevModPhys.83.1589, PhysRev.135.A550, larkin1965inhomogeneous,kopnin2001theory,Fausti2011,Kundu2013,Mankowsky2014,Khim2021}, real-world superconductors are inherently open systems, interacting with their environment. This interaction introduces dissipative effects, which are effectively captured by non-Hermitian (NH) Hamiltonians \cite{moiseyev2011non}. Despite being relatively underexplored, NH superfluid systems have recently been associated with intriguing phenomena, including $\mathcal{PT}$-symmetric quantum critical effects \cite{Ashida2017,Lourenco2018}, nonorthogonal Majorana zero modes \cite{Kawabata2018}, exceptional odd-frequency pairing \cite{Cayao2022}, the paramagnetic Meissner effect \cite{Tamura2025}, unconventional phase transitions \cite{Yamamoto2019,Yamamoto2021,Iskin2021,He2021,Li2023PRL,Shi2024,Li2024,Tajima2024,Takemori2024}, NH-modulated superconducting fluctuation \cite{Chtchelkatchev2012}, and exceptional fermionic superfluidity \cite{Takemori2025}.

Recent experimental advances in ultracold atomic systems have enabled the realization of superfluid states mediated by orbital Feshbach resonance in alkaline-earth atoms \cite{Zhang2015, He2016, Hoefer2015, Pagano2015, DarkwahOppong2019, Cappellini2019}. These systems exhibit metastable states with strong inelastic collisions, resulting in tunable two-body loss \cite{Tomita2019}. Unlike elastic scattering or one-body loss, two-body loss introduces complex-valued interaction, offering a novel platform for investigating NH superfluid.
A mean-field theory based on the biorthogonal basis has been initially developed to demonstrate how superfluidity is modified under two-body loss \cite{Yamamoto2019,Shi2024}. However, recent studies have shown that the right-eigenstate-based definition of observables is both experimentally more relevant \cite{Barontini2013,Li2019,Wu2019,Yamamoto2022,Yamamoto2023} and physically meaningful \cite{Meden2023,Graefe2008,Dora2019,Sticlet2022,Yan2024}. For instance, studies on normal metal-insulator-NH superconductor junctions have shown that the right-eigenstate-based definition accurately describes the Andreev reflection process. In contrast, using the biorthogonal basis to define current leads to a predicted reduction in conductance due to Andreev reflection, which violates fundamental physical principles. \cite{Kornich2023}. 
As we know, in the context of mean-field theories, the definition of the order parameter plays a pivotal role. A right-eigenstate-based NH mean-field theory has been developed, successfully describing the first-order $\mathcal{PT}$ phase transitions in NH superconductors \cite{gain_loss}. However, within this framework, the effects of two-body loss and impurity scattering on superfluidity remain an open question, warranting further exploration.

%figure2
%figure in one column
\begin{figure*}[hbt] 
\centering
\includegraphics[width=0.95\textwidth]{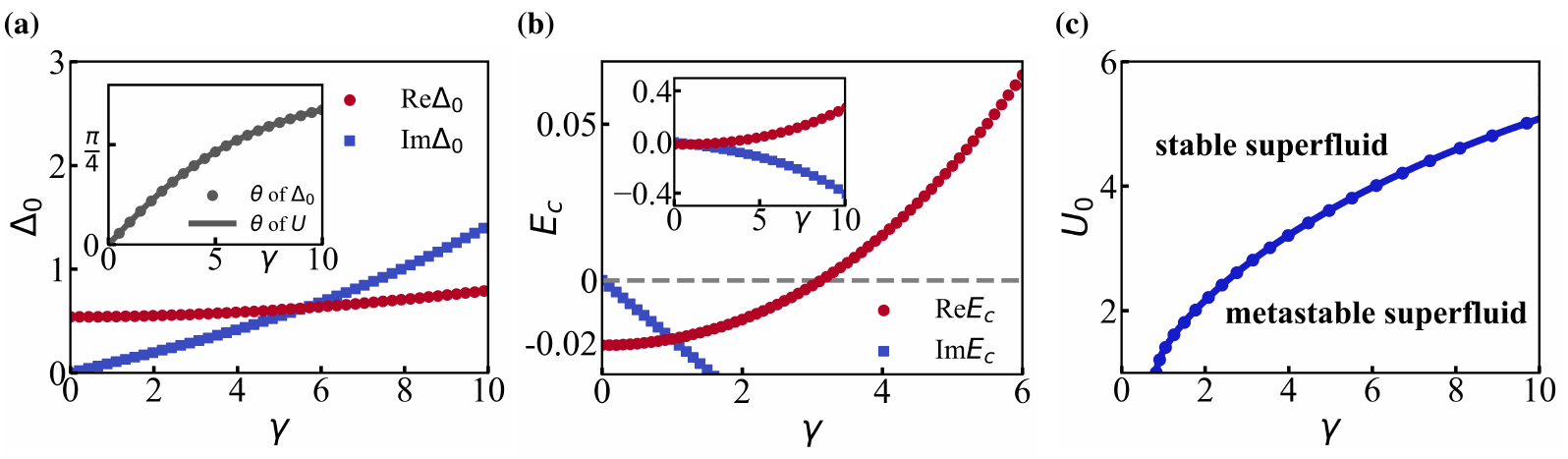} 
\caption{Numerical solutions derived from the NH gap equation [Eq.~\eqref{equ:M1:eqgapeq}]. (a) $\Delta_0$ plotted against $\gamma$ for $U_0=2.8$. The inset shows the complex angle $\theta$ of $\Delta_0$ (grey dot) and $U$ (grey line) as functions of $\gamma$ for $U_0=2.8$, demonstrating exact coincidence. (b) $E_c$ as a function of $\gamma$ for $U_0=2.8$, with the inset presenting results for $E_c$ over a broader range of $\gamma$. (c) The phase diagram of superfluid stability at half-filling, where the blue curve indicates $\text{Re}E_c = 0$.}
\label{fig:fig2}
\end{figure*}

In this work, we explore an open superfluid system modeled with complex-valued interaction. By numerically solving the gap equations derived from both the right-eigenstate-based mean-field theory and its biorthogonal counterpart, we evaluate the suitability of these two approachs in describing the NH superfluid systems. We find that the biorthogonal approach fails to yield nontrivial solutions in the regime of moderate dissipation and exhibits discontinuities. In contrast, the right-eigenstate-based approach successfully produces continuous solutions and achieves lower condensation energy, indicating its superior suitability for describing the NH superfluid systems.
Furthermore, our analysis reveals that although two-body loss amplifies the magnitude of the order parameter, it concurrently introduces a positive real part to the condensation energy, thereby driving the superfluid state into a metastable regime. In contrast, the intrinsic s-wave pairing interaction mitigates this destabilizing effect, promoting the stabilization of the superfluid state. The interplay between them gives rise to a transition in superfluid stability. Additionally, backscattering further disrupts superfluidity, causing the order parameter rapidly decay to zero. Notably, moderate backscattering amplifies the destabilizing effect of two-body loss, thereby shrinking the stable superfluid phase in the phase diagram.

% \section{Model systems}
\mytitle{Model and the NH gap equation}.--We consider an NH superfluid model on a cubic lattice with two-body loss described by the Hamiltonian
\begin{equation}
\label{equ:M1:eqH}
H=-w \sum_{\langle n, m\rangle} c_n^{\dagger} c_m-\mu \sum_n c_n^{\dagger} c_n-U \sum_n c_{n \uparrow}^{\dagger} c_{n \downarrow}^{\dagger} c_{n \downarrow} c_{n \uparrow}.
\end{equation}
Here, $w$ represents the nearest-neighbor hopping amplitude, $\mu$ is the chemical potential, and $U = U_0 + i\gamma/2$, where $\gamma$ represents the two-body loss rate.
The two-body loss, induced by the coupling between the environment and the superfluid, results in a complex-valued interaction $U$, as described by the Lindblad quantum master equation \cite{PhysRevLett.68.580, Suppl, Yamamoto2019}, which fundamentally accounts for the emergence of non-Hermiticity.

According to the right-eigenstate-based NH mean-field theory \cite{gain_loss}, the order parameters are defined as
\begin{equation}
\label{equ:M1:eqorder}
\begin{aligned}
& \Delta=-\frac{U}{N} \sum_k \frac{\left\langle\Psi_0\right| c_{-k\downarrow} c_{k\uparrow}\left|\Psi_0\right\rangle}{\left\langle\Psi_0 \mid \Psi_0\right\rangle}, \\
& \bar{\Delta}=-\frac{U}{N} \sum_k \frac{\left\langle\Psi_0\right| c_{k\uparrow}^{\dagger} c_{-k\downarrow}^{\dagger}\left|\Psi_0\right\rangle}{\left\langle\Psi_0 \mid \Psi_0\right\rangle},
\end{aligned}
\end{equation}
where $N$ is the number of unit cells, and $|\Psi_0\rangle$ denotes the NH Bardeen-Cooper-Schrieffer (BCS) ground state as detailed below \cite{PhysRev.106.162}. 
By performing the mean-field approximation, we obtain
\begin{equation}
\label{equ:M1:eqHMF}
H_{M F}=\sum_k\left(\begin{array}{ll}
c_{k \uparrow}^{\dagger} & c_{-k \downarrow}
\end{array}\right)\left(\begin{array}{cc}
\xi_k & \Delta \\
\bar{\Delta} & -\xi_k
\end{array}\right)\binom{c_{k \uparrow}}{c_{-k \downarrow}^{\dagger}},
\end{equation}
where $\xi_k=\varepsilon_k - \mu$ and $\varepsilon_k = -2 w (\cos k_x+\cos k_y+\cos k_z)$ is the energy dispersion of a cubic lattice with nearest-neighbor hopping.
Eq.~\eqref{equ:M1:eqHMF} can be diagonalized as $H_{M F}=\sum_k E_k\left(\bar{\gamma}_{k \uparrow} \gamma_{k \uparrow}+\bar{\gamma}_{-k \downarrow} \gamma_{-k \downarrow}\right)$ by the NH Bogoliubov transformation, where $\bar \gamma_{ k\sigma}, \gamma_{ k\sigma}$ are the quasiparticle creation and annihilation operators with detailed forms in the Supplemental Materials \cite{Suppl}, and $E_{k}= \sqrt {\xi_k^2 + \Delta \bar{\Delta}}$ are the quasiparticle excitation energies. 
It is noteworthy that, unlike the Hermitian case, $\bar{\gamma}_{k\sigma} \neq \gamma_{k\sigma}^{\dagger}$; these quasiparticles are neither conventional fermions nor bosons, although $\left \{\bar\gamma_{k\sigma},\gamma_{k'\sigma'}\right \} = \delta_{kk'}\delta_{\sigma\sigma'}$. 
In the ground state, all quasiparticle excitations are absent. Consequently, $|\Psi_0\rangle = \prod_{k} \gamma_{ k \uparrow} \gamma_{ -k \downarrow} |0\rangle$, where $|0\rangle$ denotes the fermionic vacuum.
Combining this with Eq.~\eqref{equ:M1:eqorder}, we derive the NH gap equation
\begin{equation}
\label{equ:M1:eqgapeq}
\frac{1}{U}=\frac{1}{2N} \sum_k \frac{E_k^*+\xi_k}{\left(\left|u_k\right|^2+\left|v_k\right|^2\right)\left|E_k\right|\left|E_k+\xi_k\right|},
\end{equation}
where $E_k^*$ denotes the complex conjugate of $E_k$, and $u_k$, $v_k$ represent the coefficients of the NH Bogoliubov transformation, with their explicit forms provided in the Supplemental Materials \cite{Suppl}.
In the cubic lattice model, both theoretical considerations and numerical calculations indicate that small changes to the Fermi level have little impact on the system's physical properties.
For simplicity, we focus on the half-filling case ($\mu = 0$) and normalize the energy scale by setting $w = 1$ in the subsequent discussion.

It is essential to emphasize that, although the interaction is complex-valued, the U(1) symmetry of the original Hamiltonian Eq.~\eqref{equ:M1:eqH} is preserved. With the mean-field treatment [Eq.~\eqref{equ:M1:eqorder}], we can generally define
\begin{equation}
\label{equ:M1:eqDphase}
\begin{aligned}
\Delta=\Delta_0 e^{i \phi}, \ \ \ \bar{\Delta}=\Delta_0 e^{-i \phi},
\end{aligned}
\end{equation}
where $\phi$ represents the U(1) gauge phase and $\Delta_0 \in \mathbb{C}$. Under this definition, $\Delta\bar{\Delta}=\Delta_0^2$ is gauge-invariant. Consequently, $E_k$ and the gap equation are also independent of the gauge phase $\phi$. Moreover, $\Delta_0$ serves as the order parameter, with both its real and imaginary components being gauge-invariant and essential to the properties of the system. 
It can be analytically and numerically demonstrated that $\text{Re}\Delta_0$ corresponds to the real part of the spectral gap, while $\text{Im}\Delta_0$ coincides with the maximum of $\text{Im}E_{k}$.
In the Hermitian limit ($\gamma = 0$), $E_{ k} = \sqrt{\xi_k^2 + \Delta_0^{2}}$, reducing Eq.~\eqref{equ:M1:eqgapeq} to $\frac{1}{U} = \frac{1}{2N} \sum_k \frac{1}{E_k}$.

% \section
\mytitle{Numerical solutions of the gap equation}.--By solving the NH gap equation [Eq.~\eqref{equ:M1:eqgapeq}] numerically, we obtain the dependence of $\Delta_0$ on the two-body loss rate $\gamma$ with a fixed $U_0=2.8$, as illustrated in Fig.~\ref{fig:fig2}(a). 
With the gauge choice $\phi=0$, $U$ and $\Delta_0$ are shown to share the same complex angle $\theta$. Meanwhile, $\text{Re}\Delta_0$ exhibits a positive correlation with $U_0$ and remains nearly invariant with respect to $\gamma$, whereas $\text{Im}\Delta_0$ is positively correlated with $\gamma$.
%In the strong dissipation regime, the quantum Zeno effect (QZE) induces particle localization, which in turn facilitates superfluid. As a result, the real part of the order parameter, $\text{Re}\Delta_0$, increases with $\gamma$.
The inset of Fig.~\ref{fig:fig2}(a) illustrates the phase-locking between $\Delta_0$ (dots) and $U$ (line).

In NH systems, energy possesses both real and imaginary components, where the real part is associated with particle occupations, while the imaginary part reflects particle lifetimes. Accordingly, the real part of the condensation energy, $\text{Re}E_c$, indicates whether the NH superfluid state is energetically favorable \cite{Yamamoto2019}.
For $\mu=0$, the particle number is given by $N_e=\sum_{k \sigma}\left\langle\Psi_0\right| c_{k \sigma}^{\dagger} c_{k \sigma}\left|\Psi_0\right\rangle /\left\langle\Psi_0 \mid \Psi_0\right\rangle=N$, corresponding to the half-filling condition. Under this scenario, the condensation energy $E_c$ is
\begin{equation}
E_c=\frac{1}{N} \sum_k\left(-E_k+\xi_k+\left|\xi_k\right|\right)+\frac{1}{U} \Delta_0^2.
\end{equation}
Fig.~\ref{fig:fig2}(b) illustrates the dependence of $E_c$ on $\gamma$ with a fixed $U_0=2.8$. As $\gamma$ increases, $\text{Re}E_c$ rises and eventually surpasses zero, indicating that the superfluid state ceases to be the stable ground state of the system.
The two-body loss $\gamma$, by virtue of the phase-locking property, facilitates the formation of the superfluid order parameter. However, the induced non-Hermiticity ultimately undermines the stability of the system, rendering the superfluid state metastable.
The inset of Fig.~\ref{fig:fig2}(b) shows $E_c$ over a wider range of $\gamma$.

Building on the analysis of $\text{Re}E_c$, we construct the superfluid stability phase diagram in the $U_0-\gamma$ plane, as shown in Fig.~\ref{fig:fig2}(c). 
Here, we only choose $U_0 > 1$, since it is computationally more demanding to reach the required accuracy for smaller $U_0$.
The transition in superfluid stability, driven by two-body loss, emerges as a universal phenomenon.
For a fixed $U_0$, as $\gamma$ increases, the system transitions from a stable superfluid state to a metastable superfluid state. 
Notably, the larger the $U_0$, the higher the critical value of $\gamma$ at the transition point.
In summary, the pairing interaction preserves superfluid stability, while two-body loss undermines it, and their competition ultimately governs the superfluid stability transitions.

%figures
%figure in two column
\begin{figure}[t]
\includegraphics[width=5.9cm]{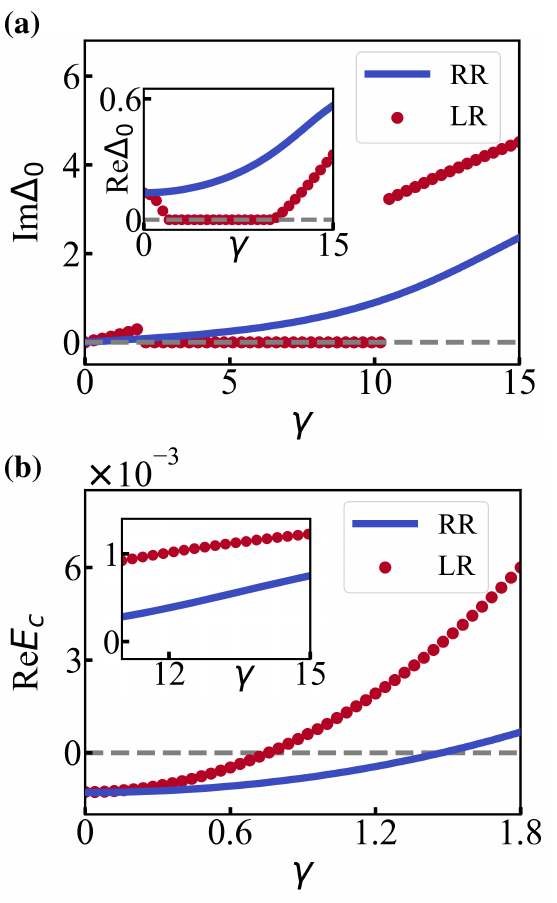}
\caption{Comparison of numerical solutions to the NH gap equation under different definitions of the order parameters. (a) $\Delta_0$ and (b) $\text{Re}E_c$ as functions of $\gamma$ for $U_0=1.8$. The blue lines represent solutions based on the right-eigenstate-based approach (RR), while the red points correspond to solutions based on the biorthogonal approach (LR). The inset in (b) highlights the results for the metastable superfluid phase, whereas the main panel depicts the stable superfluid phase.}
\label{fig:figs}
\end{figure}

% \section
\mytitle{The two definitions of the order parameters}.--In NH physics, the eigenstates of an NH Hamiltonian are generally non-orthogonal, and their evolution is inherently nonunitary \cite{Brody2013}. These characteristics introduce unique challenges and stimulate discussions regarding the proper definition of observables.
For an NH matrix $O$ of dimension $n\times n$ with $n$ nondegenerate eigenvalues, or with the degenerate eigenvalues where the algebraic multiplicity equals the geometric multiplicity, its right eigenstates constitute a complete basis \cite{Suppl}.
At present, two primary approaches to defining observables exist: one based on the right eigenstate and the other on the biorthogonal basis \cite{Meden2023}.
While the biorthogonal definition provides a more concise mathematical framework, the right-eigenstate-based definition offers a more intuitive probabilistic interpretation \cite{Meden2023,Graefe2008,Dora2019,Sticlet2022,Yan2024,Kornich2023} and aligns more closely with experimental observations \cite{Barontini2013,Li2019,Wu2019,Yamamoto2022,Yamamoto2023}.
The choice of order parameter definition is particularly crucial in the context of mean-field theory.
In this section, we investigate the applicability of these two definitions to the NH superfluid systems, based on numerical simulation results from the complex-valued interaction superfluid model.

The order parameter $\Delta_{0}^{LR}$, defined within the biorthogonal basis, satisfies the gap equation $\frac{1}{U} = \frac{1}{2N} \sum_k \frac{1}{E_k^{LR}}$, where $E_{k}^{LR} = \sqrt{\xi_k^2 + (\Delta_{0}^{LR})^2}$ \cite{Suppl}. With a fixed $U_0=1.8$, we numerically calculate $\Delta_{0}$ and $\text{Re}E_c$ as functions of $\gamma$ under both frameworks.
Firstly, as illustrated in Fig.~\ref{fig:figs}(a), within the moderate dissipation region, the biorthogonal approach yields no nontrivial solutions (red points), indicating a normal phase. As $\gamma$ increases further, $\Delta_{0}^{LR}$ undergoes an abrupt change, leading to the reemergence of superfluidity. 
 However, these abnormal discontinuities can be readily avoided by using the right-eigenstate-based approach. As shown in Fig.~\ref{fig:figs}(a), blue solid lines are shown as continuous solutions across the entire parameter space, without any nonanalytic behavior.
Secondly, as shown in Fig.~\ref{fig:figs}(b), the right-eigenstate-based approach results in lower condensation energy, applicable to both stable and metastable superfluid phases.
We thereby conclude that, for complex-valued interaction superfluid, the right-eigenstate-based mean-field theory proves to be more suitable. This finding further supports discussions in related literature \cite{gain_loss}.

% \section
\mytitle{The backscattering effect}.--In this section, we examine the impact of impurity scattering on the NH superfluid using an exactly solvable simplified model—the backscattering model. This model posits that electrons with momentum $k$, upon encountering an impurity, have their momentum reversed to $-k$. Although this simplified model does not precisely represent real random impurities or phonon scattering, it effectively captures the attenuation of flow and the finite lifetime of quasiparticles induced by impurity scattering in many-body systems \cite{PhysRevLett.71.2304}.

The NH superfluid model with two-body loss in the presence of backscattering is described by
\begin{equation}
\label{equ:M1:eqHBS}
\begin{aligned}
H_{M F}^{B S}&=H_{M F}+\sum_{k>0, \sigma} \Gamma_{BS}\left(c_{k \sigma}^{\dagger} c_{-k \sigma}+c_{-k \sigma}^{\dagger} c_{k \sigma}\right),
\end{aligned}
\end{equation}
where $\Gamma_{BS}$ represents the backscattering strength and $H_{MF}$ is detailed in Eq.~\eqref{equ:M1:eqHMF}. $H_{M F}^{B S}$ can be diagonalized as
$H_{M F}^{B S}=\sum_{k>0} E_{\alpha k}\left(\bar{\gamma}_{\alpha k \uparrow} \gamma_{\alpha k \uparrow}+\bar{\gamma}_{\alpha k \downarrow} \gamma_{\alpha k \downarrow}\right)$ through an NH Bogoliubov transformation \cite{Suppl},
where $\alpha =1,2$ labels the branches of quasiparticles and $ \bar{\gamma}_{\alpha k \sigma},\gamma_{\alpha k \sigma}$ are quasiparticle operators. The quasiparticle excitation energy is given by $E_{\alpha k}=\sqrt{\left[\xi_k-(-1)^\alpha \Gamma_{BS}\right]^2+\Delta \bar{\Delta}}$.
Thus, the NH BCS ground state is defined as $\left|\Psi_0\right\rangle_{BS}=\prod_{\alpha, k>0} \gamma_{\alpha k \uparrow} \gamma_{\alpha k \downarrow}|0\rangle$. 
Combining this with Eq.~\eqref{equ:M1:eqorder}, we derive the NH gap equation in the presence of backscattering
\begin{equation}
\label{equ:M1:eqgapeqBS}
\frac{1}{U}=\frac{1}{2N} \sum_{\alpha, k>0}\left \{\frac{E_{\alpha k}^*+\left [\xi_k-(-1)^\alpha\Gamma_{BS}\right]}{C\left|E_{\alpha k}\right|\left|E_{\alpha k}+\left [\xi_k-(-1)^\alpha\Gamma_{BS}\right]\right|}\right \},
\end{equation}
where $C$ is the normalization factor of $\left|\Psi_0\right\rangle_{BS}$ \cite{Suppl}. In the weak scattering limit ($\Gamma_{BS} \to 0$), this equation simplifies to Eq.~\eqref{equ:M1:eqgapeq}.  
Furthermore, in the Hermitian limit ($\gamma =0$), it reduces to $\frac{1}{U}= \frac{1}{2 N} \sum_{\alpha, k>0} \frac{1}{E_{\alpha k}}$ where $E_{\alpha k}=\sqrt{\left[\xi_k-(-1)^\alpha \Gamma_{BS}\right]^2+\Delta_0^2}$ \cite{PhysRevLett.71.2304}.
It is crucial to emphasize that, with the definition of the order parameters [Eq.~\eqref{equ:M1:eqorder}] being unaffected by the backscattering, the formulation in Eq.~\eqref{equ:M1:eqDphase} remains applicable. As a result, $\Delta_0$, $E_{\alpha k}$ and the gap equation all exhibit gauge invariance.

%figure3
\begin{figure}[t]
\includegraphics[width=5.7cm]{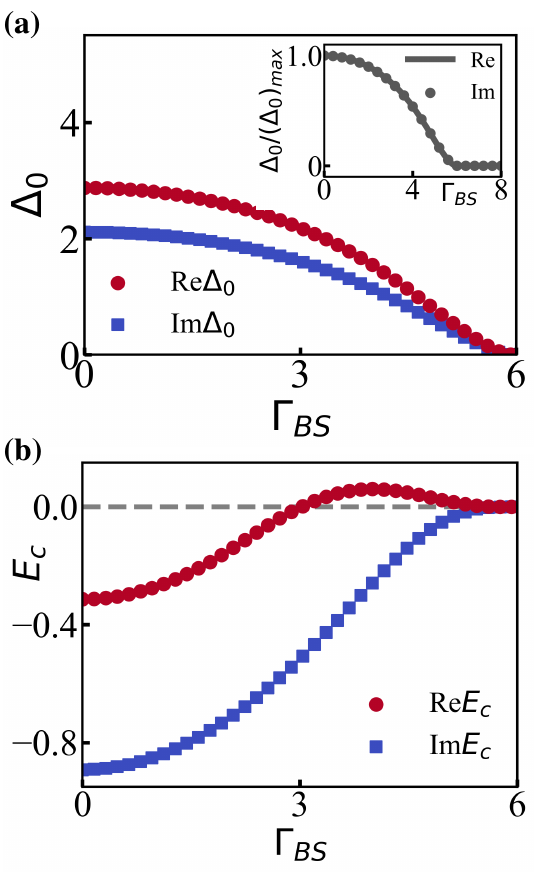}
\caption{Numerical solutions derived from the NH gap equation with backscattering [Eq.~\eqref{equ:M1:eqgapeqBS}]. $\Delta_0$ (a) and $E_c$ (b) are shown as functions of $\Gamma_{BS}$ for $U_0=6.8$ and $\gamma=10$. The inset in (a) displays $\text{Re}\Delta_0$ (grey line) and $\text{Im}\Delta_0$ (grey dot) normalized to their maximum values. }
\label{fig:fig3}
\end{figure}

By the NH gap equation in the presence of backscattering [Eq.~\eqref{equ:M1:eqgapeqBS}], we analyze the dependence of $\Delta_0$ on the backscattering strength $\Gamma_{BS}$ for $U_0 = 6.8$ and $\gamma = 10$, as shown in Fig.~\ref{fig:fig3}(a). With increasing $\Gamma_{BS}$, $\Delta_0$ rapidly decays to $0$, driving the system into a normal state where stable Cooper pairs can no longer form, similar to the Hermitian case \cite{PhysRevLett.71.2304}.
Upon normalization to the maximum value, both $\text{Re}\Delta_0$ and $\text{Im}\Delta_0$ exhibit identical dependence on $\Gamma_{BS}$, as shown in the inset of Fig.~\ref{fig:fig3}(a). This behavior stems from the phase-locking between $\Delta_0$ and $U$, which ensures that the phase of $\Delta_0$ remains invariant with respect to $\Gamma_{BS}$. Consequently, for a fixed $U$, as $\Gamma_{BS}$ varies, the relationship $\operatorname{Re} \Delta_0 / \operatorname{Re} \Delta_{0 \max } = \operatorname{Im} \Delta_0 / \operatorname{Im} \Delta_{0 \text{max}} = \left|\Delta_0\right| / \left|\Delta_0\right|_{\max}$ is maintained.

By calculating the particle number for $\mu=0$, we find $N_e=N$, indicating the half-filling. Thus, the condensation energy $E_c$ is expressed as
\begin{equation}
E_c = \frac{1}{N} \sum_{\alpha, k} \left( -\frac{1}{2} E_{\alpha k} + \xi_k + \left| \xi_k - |\Gamma_{BS}| \right| \right) + \frac{1}{U} \Delta_0^2.
\end{equation}
Fig.~\ref{fig:fig3}(b) demonstrates the relationship between $E_c$ and $\Gamma_{BS}$ with $U_0 = 6.8$ and $\gamma = 10$. The results reveal that, facilitated by two-body loss, moderate backscattering can destabilize the superfluid state, thereby inducing a superfluid stability transition. Specifically, when backscattering is initially introduced, $\text{Re}E_c<0$, signifying a stable superfluid phase. As $\Gamma_{BS}$ increases, the system transitions to a metastable superfluid state, characterized by $\text{Re}E_c>0$. Finally, with further increases in $\Gamma_{BS}$, the superfluid state is entirely destroyed, and the system transitions into the normal state, where $\text{Re}E_c=0$.

%figure4
\begin{figure}[t]
\includegraphics[width=5.7cm]{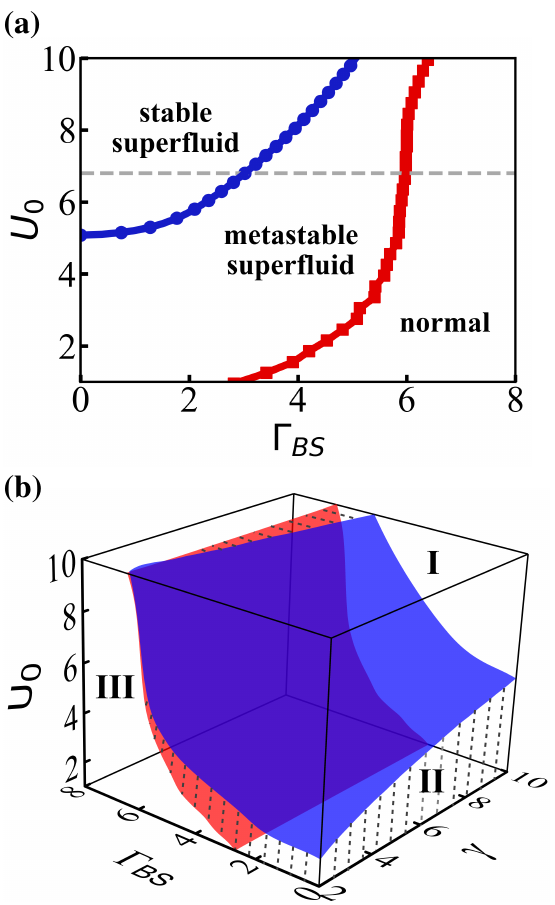}
\caption{Phase diagrams. (a) Phase diagram in the $U_0$-$\Gamma_{BS}$ plane at $\gamma=10$. The grey dashed line corresponds to the phase transition described in Fig.~\ref{fig:fig3}; (b) Three-dimensional phase diagram in the $U_0$-$\gamma$-$\Gamma_{BS}$ parameter space. Region I: the stable superfluid phase; Region II: the metastable superfluid phase; Region III: the normal phase. The blue curve (or surface) represents the condition $\text{Re}E_c = 0$ with $\Delta_0 \ne 0$, while the red curve (or surface) signifies the onset of $\Delta_0 = 0$.}
\label{fig:fig4}
\end{figure}

By examining the order parameter and condensation energy, we delineate the phase diagram on the $U_0$-$\Gamma_{BS}$ plane with $\gamma=10$, revealing the complex interplay between paring interaction, two-body loss, and backscattering, as shown in Fig.~\ref{fig:fig4}(a).
In the absence of backscattering ($\Gamma_{BS}=0$), two-body loss destabilizes the superfluid stability, while the pairing interaction $U_0$ preserves it. Their competition induces a superfluid stability transition. 
When backscattering is introduced, $U_0$ required to sustain a stable superfluid phase increases significantly, as indicated by the blue curve in Fig.~\ref{fig:fig4}(a). This observation suggests that backscattering amplifies the destabilizing effect of two-body loss on superfluidity. 
With further increase in $\Gamma_{BS}$, the system transitions into the normal phase with both $\Delta_0=0$ and $\text{Re}E_c=0$, and the transition curve is marked red, as shown in Fig.~\ref{fig:fig4}(a). Notably, at the critical point, a larger $U_0$ corresponds to a higher $\Gamma_{BS}$.
Fig.~\ref{fig:fig4}(b) presents the three-dimensional phase diagram within the $U_0$-$\gamma$-$\Gamma_{BS}$ parameter space. The transition to the normal phase, represented by the red surface, is exclusively governed by $\Gamma_{BS}$. As shown in Fig.~\ref{fig:figs1}(d) in the Supplemental Materials, the cross-sections of the red surface for various values of $\gamma$ exhibit complete overlap \cite{Suppl}. In contrast, the superfluid stability transition, depicted by the blue surface, is jointly influenced by both $\gamma$ and $\Gamma_{BS}$. Within the superfluid phase, maintaining stability becomes increasingly challenging as the effects of two-body loss and backscattering intensify.

In short, the pairing interaction strength $U_0$ promotes the superfluid phase formation, whereas two-body loss and backscattering act as destabilizing forces. The interplay among these factors leads to the emergence of a distinctive metastable superfluid phase in NH superfluid systems.
Currently, fermionic superfluidity under two-body loss has been experimentally achieved in ultracold atomic systems, such as ytterbium atoms \cite{DarkwahOppong2019, Tomita2019}. In these systems, the interaction strength is tunable via orbital Feshbach resonance \cite{Zhang2015, He2016, Hoefer2015, Pagano2015, Cappellini2019}, while the two-body loss rate can be precisely controlled using photoassociation techniques \cite{Tomita2017}. The impact of two-body loss on the superfluid gap can be experimentally probed through angle-resolved photoemission spectroscopy \cite{Damascelli2003, Vishik2010, DarkwahOppong2019, Brown2019}. These advancements in ultracold atomic systems establish a versatile and robust platform for investigating both stable and metastable non-Hermitian superfluid phases, hopefully will provide valuable insights into the unconventional physics in these systems.

\mytitle{Summary.}--We have systematically investigated an NH superfluid model with complex-valued interaction by employing a right-eigenstate-based mean-field theory. Firstly, our work resolves a critical ambiguity in the theoretical description of such systems, demonstrating that the right-eigenstate-based approach provides a more physically consistent and energetically favorable framework than its biorthogonal counterpart. Secondly, we identified a mechanism where two-body loss, while enhancing the pairing amplitude, paradoxically drives the system towards a metastable state. Thirdly, we showed that backscattering acts as a potent destabilizing agent that can ultimately destroy the superfluid phase. These results underscore the delicate interplay between pairing, dissipation, and disorder in NH superfluids. This work not only provides a robust theoretical tool for analyzing open superfluids but also offers direct, testable predictions for ultracold atom experiments with tunable two-body loss. Future investigations could extend this framework to explore non-equilibrium dynamics, the role of thermal fluctuations, and the potential for engineering novel, dissipation-stabilized topological phases in NH systems.

\let\oldaddcontentsline\addcontentsline% Store \addcontentsline
\renewcommand{\addcontentsline}[3]{}% Make \addcontentsline a no-op

\mytitle{Acknowledgements.}--This work was financially supported by the National Key Research and Development Program of China (Grant No. 2024YFA1409001), the National Natural Science Foundation of China (Grants No.~12374037 and No.~12204044), and the Strategic Priority Research Program of the Chinese Academy of Sciences (Grant No. XDB28000000), and the Fundamental Research Funds for the Central Universities.

\bibliography{reference.bib}  % The references (bibliography) information are stored in the file named "Bibliography.bib"
\let\addcontentsline\oldaddcontentsline% Restore \addcontentsline

\newpage
\onecolumngrid
\newpage
{
	\center \bf \large 
	Supplemental Materials\\
	%\large for ``Non-Hermitian exceptional Landau quantization in electric circuits"\vspace*{0.1cm}\\ 
	\large for ``\newtitle"\vspace*{0.1cm}\\ 
	\vspace*{0.5cm}
	%\newauthor
}

\tableofcontents

\setcounter{equation}{0}
\setcounter{figure}{0}
\setcounter{table}{0}
\setcounter{page}{1}

\renewcommand{\theequation}{S\arabic{equation}}
\renewcommand{\thefigure}{S\arabic{figure}}
\renewcommand{\theHtable}{Supplement.\thetable}
\renewcommand{\theHfigure}{Supplement.\thefigure}
\renewcommand{\bibnumfmt}[1]{[S#1]}

% section
\section{The right-eigenstate-based non-Hermitian mean-field theory}
\subsection{The non-Hermitian superfluid with two-body loss}
\label{subsec:two_body_loss}
The evolution of an open quantum system is governed by the Lindblad quantum master equation
\begin{equation}
\label{equ:S1:eqlind}
\begin{aligned}
\frac{d\rho_s(t)}{dt} & = -i[H_0, \rho_s(t)] - \frac{1}{2} \gamma \sum_n \left[ L_n^{\dagger} L_n \rho_s(t) + \rho_s(t) L_n^{\dagger} L_n - 2 L_n \rho_s(t) L_n^{\dagger} \right],
\end{aligned}
\end{equation}
where $\rho_s$ denotes the reduced density matrix, and $L_n$ represents the Lindblad operator characterizing the loss at site $n$ with strength $\gamma$ \cite{PhysRevLett.68.580}. When the timescale of dynamical evolution is much shorter than $1/\gamma$, the quantum jump effects encapsulated by the last term in Eq.~\eqref{equ:S1:eqlind} can be neglected. In this case, the coupling with the environment can be effectively described by a non-Hermitian (NH) Hamiltonian
\begin{equation}
H_{\mathrm{eff}}=H_0-\frac{i}{2} \gamma \sum_n L_n^{\dagger} L_n.
\end{equation}
For atomic gases experiencing two-body losses due to inelastic collisions, the Lindblad operators are given by $L_n = c_{n \downarrow} c_{n \uparrow}$. Thus, the NH superfluid model with two-body loss can be effectively described by
\begin{equation}
\label{equ:S1:eqH}
H=-w \sum_{\langle n, m\rangle} c_n^{\dagger} c_m-\mu \sum_n c_n^{\dagger} c_n-U \sum_n c_{n \uparrow}^{\dagger} c_{n \downarrow}^{\dagger} c_{n \downarrow} c_{n \uparrow},
\end{equation}
where $U = U_0 + i \gamma/2$, with $U_0, \gamma > 0$. 

In NH quantum mechanics, the time evolution of a quantum state $\psi(t)$ is nonunitary. This necessitates careful consideration in defining observables.
For a NH matrix $H$ of dimension $n*n$ with $n$ nondegenerate eigenvalues, $H$ possesses $n$ linearly independent right eigenstates. In cases involving degenerate eigenvalues, as long as the algebraic multiplicity equals the geometric multiplicity, the right eigenstates still form a complete basis.
Therefore, the expectation value of operator $O$ in the state $|\psi\rangle$ can be defined as 
\begin{equation}
\label{equ:A1:eqd10}
\langle O\rangle=\frac{\langle\psi| O|\psi\rangle}{\langle\psi \mid \psi\rangle} ,
\end{equation}
where the denominator is proper normalization, maintaining the conservation of probability\cite{Sticlet2022}. This definition follows the ideas from the Hermitian quantum mechanics and is both experimentally more relevant \cite{Barontini2013,Li2019,Wu2019} and physically meaningful \cite{Meden2023,Sticlet2022,Yan2024,Dora2019,Yamamoto2021,Yamamoto2022}. We refer to this method as the right-eigenstate-based definition. 
Another framework for defining observables in NH systems relies on the biorthogonal basis \cite{Brody2013,Yamamoto2019}. While the eigenvalues of an NH Hamiltonian $H$ can be accurately obtained through the biorthogonal basis, it is crucial to emphasize that these eigenvalues are determined solely by the eigenvalue equations, which depend exclusively on the right eigenstates. Thus, the physical properties of NH systems can be fully characterized using the right eigenstates alone. 
The left eigenstates in the biorthogonal basis serve only as auxiliary mathematical constructs for the eigenvalue decomposition of the NH matrix and bear no direct relevance to the physical description of the system. By definition, these left eigenstates are eigenstates of the adjoint matrix $H^\dagger$ and are fundamentally distinct from the right eigenstates. For instance, in a purely lossy system described by H, the adjoint operator $H^\dagger$ and its eigenstates (which are the left eigenstates of $H$) describe a purely gain system, which is physically irrelevant to the system under consideration.
Recent studies have highlighted the importance of adopting a right-eigenstate-based approach for analyzing NH systems, as naive applications of the biorthogonal basis often result in unphysical predictions. For instance, research on normal metal-insulator-NH superconductor junctions has revealed that current defining using the biorthogonal basis predicts a reduction in conductance due to Andreev reflection, which contradicts fundamental physical principles. Conversely, the right-eigenstate-based definition correctly captures the Andreev reflection process \cite{Kornich2023}.
Furthermore, from the perspective of the time-dependent Schrödinger equation, the time evolution of right and left eigenstates is governed by $H$ and $H^\dagger$, respectively. In time-dependent dynamical problems, the inconsistency of defining observables based on biorthogonal basis becomes even more pronounced.

As we know, the definition of the order parameter is crucial in the mean-field theories. In this context, a right-eigenstate-based NH mean-field theory has been formulated, successfully capturing the first-order $\mathcal{PT}$ phase transitions in NH superconductors \cite{gain_loss}.
In the right-eigenstate-based NH mean-field theory, the order parameters of NH superfluid with two-body loss are defined as
\begin{equation}
\label{equ:S1:eqorder}
\begin{aligned}
& \Delta=-\frac{U}{N} \sum_k\left\langle c_{-k \downarrow} c_{k\uparrow} \right\rangle \equiv -\frac{U}{N} \sum_k \frac{\left\langle\Psi_0\right| c_{-k\downarrow} c_{k\uparrow}\left|\Psi_0\right\rangle}{\left\langle\Psi_0 \mid \Psi_0\right\rangle}, \\
& \bar{\Delta}=-\frac{U}{N} \sum_k\left\langle c_{k\uparrow}^{\dagger} c_{-k\downarrow}^{\dagger}\right\rangle \equiv -\frac{U}{N} \sum_k \frac{\left\langle\Psi_0\right| c_{k\uparrow}^{\dagger} c_{-k\downarrow}^{\dagger}\left|\Psi_0\right\rangle}{\left\langle\Psi_0 \mid \Psi_0\right\rangle},
\end{aligned}
\end{equation}
where $|\Psi_0\rangle$ is the NH Bardeen-Cooper-Schrieffer (BCS) ground state \cite{PhysRev.106.162}, as detailed below. It is important to note that, unlike in the Hermitian case, $\bar{\Delta} \ne \Delta^*$ in the current case; however, they satisfy $\Delta^*/ \bar{\Delta}=U^*/U$.
Substituting $c_{-k \downarrow} c_{k \uparrow}=\left\langle c_{-k \downarrow} c_{k \uparrow}\right\rangle+\delta\left(c_{-k \downarrow} c_{k \uparrow}\right)$ and $c_{k\uparrow}^{\dagger} c_{-k\downarrow}^{\dagger} =\left\langle c_{k\uparrow}^{\dagger} c_{-k\downarrow}^{\dagger} \right\rangle + \delta\left(c_{k\uparrow}^{\dagger} c_{-k\downarrow}^{\dagger}\right)$ into the pairing interaction term and neglecting the second-order terms of $\delta$, we obtain the NH mean-field Hamiltonian
\begin{equation}
\label{equ:S1:eqHMF}
H_{M F}=\sum_k\left(\begin{array}{ll}
c_{k \uparrow}^{\dagger} & c_{-k \downarrow}
\end{array}\right)\left(\begin{array}{cc}
\xi_k & \Delta \\
\bar{\Delta} & -\xi_k
\end{array}\right)\binom{c_{k \uparrow}}{c_{-k \downarrow}^{\dagger}}+\sum_k \xi_k+\frac{N}{U} \Delta \bar{\Delta},
\end{equation}
where $\xi_k=\varepsilon_k - \mu$ and $\varepsilon_k = -2 w (\cos k_x+\cos k_y+\cos k_z)$ is the energy dispersion of a cubic lattice with nearest-neighbor hopping.

Owing to the NH nature of the system, the NH mean-field Hamiltonian [Eq.~\eqref{equ:S1:eqHMF}] cannot be diagonalized via a conventional unitary transformation. To address this challenge, we employ the NH Bogoliubov transformation, which is explicitly defined as follows:
\begin{equation}
\label{equ:S1:eqBogoliubov}
\begin{aligned}
& \bar{\gamma}_{k \uparrow}=u_k c_{k \uparrow}^{\dagger}-\bar{v}_k c_{-k \downarrow} , \  \bar{\gamma}_{-k \downarrow}=\bar{v}_k c_{k \uparrow}+u_k c_{-k \downarrow}^{\dagger} , \\
& \gamma_{k \uparrow}=u_k c_{k \uparrow}-v_k c_{-k \downarrow}^{\dagger}, \  \gamma_{-k \downarrow}=v_k c_{k \uparrow}^{\dagger}+u_k c_{-k \downarrow} .
\end{aligned} 
\end{equation}
The transformation coefficients are given by
\begin{equation}
u_k=\sqrt{\frac{E_k+\xi_k}{2 E_k}}, \ \  v_k=-\sqrt{\frac{E_k-\xi_k}{2 E_k}} \sqrt{\frac{\Delta}{\bar{\Delta}}},\ \  \bar{v}_k=-\sqrt{\frac{E_k-\xi_k}{2 E_k}} \sqrt{\frac{\bar{\Delta}}{\Delta}},
\end{equation}
which satisfy $u_{k}^{2} + v_k \bar{v}_k = 1$ and $E_{k}= \sqrt {\xi_k^2 + \Delta \bar{\Delta}}$ are the quasiparticle excitation energies.

Upon performing the NH Bogoliubov transformation, Eq.~\eqref{equ:S1:eqHMF} is diagonalized as
\begin{equation}
H_{M F}=\sum_k E_k\left(\bar{\gamma}_{k \uparrow} \gamma_{k \uparrow}+\bar{\gamma}_{-k \downarrow} \gamma_{-k \downarrow}\right)+\sum_k\left(-E_k+\xi_k\right)+\frac{N}{U} \Delta \bar{\Delta},
\end{equation}
where $\bar \gamma_{ k\sigma}$ and $\gamma_{ k\sigma}$ are the creation and annihilation operator of quasiparticles.
Unlike the Hermitian case, $\bar\gamma_{ k\sigma} \ne \gamma_{ k\sigma}^{\dagger} $ in current case. Therefore, although they satisfy $\left \{\bar\gamma_{k\sigma},\gamma_{k'\sigma'}\right \} = \delta_{kk'}\delta_{\sigma\sigma'}$, these quasiparticles are neither conventional fermions nor bosons.

In the ground state, all quasiparticle excitations are absent.
Consequently, the ground state is expressed as $|\Psi_0\rangle = \prod_{k} \gamma_{ k \uparrow} \gamma_{ -k \downarrow} |0\rangle$, where $|0\rangle$ denotes the fermionic vacuum.
Utilizing Eq.~\eqref{equ:S1:eqBogoliubov}, $|\Psi_0\rangle $ can be reformulated as $\left|\Psi_0\right\rangle=\prod_k\left(u_k+v_k c_{k \uparrow}^{\dagger} c_{-k \downarrow}^{\dagger}\right)|0\rangle$, with the normalization factor $\left\langle\Psi_0 \mid \Psi_0\right\rangle=\prod_k\left(\left|u_k\right|^2+\left|v_k\right|^2\right)$. In the Hermitian limit, $\left|u_k\right|^2+\left|v_k\right|^2 = 1$.
By combining this with Eq.~\eqref{equ:S1:eqorder}, we derive the NH gap equation
\begin{equation}
\label{equ:S1:eqgapeq}
\frac{1}{U}=\frac{1}{2N} \sum_k \frac{E_k^*+\xi_k}{\left(\left|u_k\right|^2+\left|v_k\right|^2\right)\left|E_k\right|\left|E_k+\xi_k\right|}.
\end{equation}
It is important to emphasize that the gap equations derived from $\bar{\Delta}$ and $\Delta$ are complex conjugates of each other and are therefore equivalent.

Considering the U(1) symmetry of the system's Hamiltonian [Eq.~\eqref{equ:S1:eqH}], and according to the mean-field treatment in Eq.~\eqref{equ:S1:eqorder}, although $\bar{\Delta} \ne \Delta^*$, we can generally define
\begin{equation}
\label{equ:S1:eqorderphase}
\begin{aligned}
\Delta=\Delta_0 e^{i \phi}, \ \ \ \bar{\Delta}=\Delta_0 e^{-i \phi},
\end{aligned}
\end{equation}
where $\Delta_0 \in \mathbb{C}$, $\Delta$ and $\bar{\Delta}$ remain independent of each other. 
Under this definition, $\Delta\bar{\Delta}=\Delta_0^2$ is gauge-invariant. It follows naturally that $E_k$ and the gap equation [Eq.~\eqref{equ:S1:eqgapeq}] are also independent of the choice of the gauge phase $\phi$.

In the Hermitian limit ($\gamma = 0 $), where $\bar{\Delta}=\Delta^*$, $\Delta_0  \in \mathbb{R}$ and $E_{ k} = \sqrt{\xi_k^2 + \Delta_{0}^{2}}$, Eq.~\eqref{equ:S1:eqgapeq} thus simplifies to the Hermitian case
\begin{equation}
 \frac{1}{U}= \frac{1}{2 N} \sum_k \frac{1}{E_k}.
\end{equation}

\subsection{The non-Hermitian superfluid with two-body loss and backscattering}
The NH superfluid with two-body loss and backscattering is governed by the mean-field Hamiltonian \cite{PhysRevLett.71.2304}
\begin{equation}
\label{equ:S1:eqHBS}
H_{M F}^{B S}=\sum_{k \sigma} \xi_k c_{k \sigma}^{\dagger} c_{k \sigma}+\sum_k\left(\Delta c_{k \uparrow}^{\dagger} c_{-k \downarrow}^{\dagger}+\bar{\Delta}c_{-k \downarrow} c_{k \uparrow}\right)+\sum_{k>0, \sigma} \Gamma_{BS}\left(c_{k \sigma}^{\dagger} c_{-k \sigma}+c_{-k \sigma}^{\dagger} c_{k \sigma}\right),
\end{equation}
where $\Gamma_{BS}$ represents the backscattering strength, while $\Delta$ and $\bar{\Delta}$ denote the order parameters defined within the right-eigenstate-based framework, as specified in Eq.~\eqref{equ:S1:eqorder}. 

The NH mean-field Hamiltonian $H_{M F}^{B S}$ can be diagonalized using an NH Bogoliubov transformation
\begin{equation}
\label{equ:S1:eqBOBS}
\begin{aligned}
& \gamma_{\alpha k \uparrow}=\frac{1}{\sqrt{2}}\left[u_{\alpha k} c_{k \uparrow}-(-1)^\alpha u_{\alpha k} c_{-k \uparrow}-v_{\alpha k} c_{k \downarrow}^{\dagger}+(-1)^\alpha v_{\alpha k} c_{-k \downarrow}^{\dagger}\right], \\
& \bar{\gamma}_{\alpha k \uparrow}=\frac{1}{\sqrt{2}}\left[u_{\alpha k} c_{k \uparrow}^{\dagger}-(-1)^\alpha u_{\alpha k} c_{-k \uparrow}^{\dagger}-\bar{v}_{\alpha k} c_{k \downarrow}+(-1)^\alpha \bar{v}_{\alpha k} c_{-k \downarrow}\right], \\
& \gamma_{\alpha k \downarrow}=\frac{1}{\sqrt{2}}\left[v_{\alpha k} c_{k \uparrow}^{\dagger}-(-1)^\alpha v_{\alpha k} c_{-k \uparrow}^{\dagger}+u_{\alpha k} c_{k \downarrow}-(-1)^\alpha u_{\alpha k} c_{-k \downarrow}\right], \\
& \bar{\gamma}_{\alpha k \downarrow}=\frac{1}{\sqrt{2}}\left[\bar{v}_{\alpha k} c_{k \uparrow}-(-1)^\alpha \bar{v}_{\alpha k} c_{-k \uparrow}+u_{\alpha k} c_{k \downarrow}^{\dagger}-(-1)^\alpha u_{\alpha k} c_{-k \downarrow}^{\dagger}\right],
\end{aligned}
\end{equation}
which leads to
\begin{equation}
H_{M F}^{B S}=\sum_{k>0} E_{\alpha k}\left(\bar{\gamma}_{\alpha k \uparrow} \gamma_{\alpha k \uparrow}+\bar{\gamma}_{\alpha k \downarrow} \gamma_{\alpha k \downarrow}\right),
\end{equation}
where $E_{\alpha k}=\sqrt{\left[\xi_k-(-1)^\alpha \Gamma_{BS}\right]^2+\Delta \bar{\Delta}}$ are the quasiparticle excitation energies, $\alpha =1,2$ labels the branches of quasiparticles and $ \bar{\gamma}_{\alpha k \sigma},\gamma_{\alpha k \sigma}$ are the creation and annihilation operators of quasiparticles.
The transformation coefficients are given by
\begin{equation}
\begin{aligned}
& u_{\alpha k}=\sqrt{\frac{E_{\alpha k}+\left(\xi_k-(-1)^\alpha \Gamma_{BS}\right)}{2 E_{\alpha k}}}, \\
&v_{\alpha k}=-\sqrt{\frac{E_{\alpha k}-\left(\xi_k-(-1)^\alpha \Gamma_{BS}\right)}{2 E_{\alpha k}}} \sqrt{\frac{\Delta}{\bar{\Delta}}}, \\
&\bar{v}_{\alpha k}=-\sqrt{\frac{E_{\alpha k}-\left(\xi_k-(-1)^\alpha \Gamma_{BS}\right)}{2 E_{\alpha k}}} \sqrt{\frac{\bar{\Delta}}{\Delta}},
\end{aligned}
\end{equation}
satisfying $u_{\alpha k}^{2} + v_{\alpha k} \bar{v}_{\alpha k} = 1$. 

Based on this, the NH BCS ground state reads $\left|\Psi_0\right\rangle_{BS}=\prod_{\alpha, k>0} \gamma_{\alpha k \uparrow} \gamma_{\alpha k \downarrow}|0\rangle$. Utilizing Eq.~\eqref{equ:S1:eqBOBS}, $|\Psi_0\rangle_{BS} $ can be explicitly expressed as
\begin{equation}
\begin{aligned}
\left|\Psi_0\right\rangle_{BS}=\prod_{k>0} & {\left[-u_{1 k} u_{2 k}|0\rangle+\frac{1}{2}\left(v_{1 k} u_{2 k}-u_{1 k} v_{2 k}\right)\left(c_{k \downarrow}^{\dagger} c_{-k \uparrow}^{\dagger}+c_{-k \downarrow}^{\dagger} c_{k \uparrow}^{\dagger}\right)|0\rangle\right.} \\
& \left.+\frac{1}{2}\left(v_{1 k} u_{2 k}+u_{1 k} v_{2 k}\right)\left(c_{k \downarrow}^{\dagger} c_{k \uparrow}^{\dagger}+c_{k \downarrow}^{\dagger} c_{k \uparrow}^{\dagger}\right)|0\rangle-v_{1 k} v_{2 k} c_{k \downarrow}^{\dagger} c_{-k \downarrow}^{\dagger} c_{-k \uparrow}^{\dagger} c_{k \uparrow}^{\dagger}|0\rangle\right],
\end{aligned}
\end{equation}
with the normalization factor $_{BS}\left\langle\Psi_0 \mid \Psi_0\right\rangle_{BS}=\prod_{\alpha, k>0}\left(\left|u_{\alpha k}\right|^2+\left|v_{\alpha k}\right|^2\right)$. 
By combining this with Eq.~\eqref{equ:S1:eqorder}, we derive the NH gap equation in the presence of backscattering
\begin{equation}
\label{equ:S1:eqBSgapeq}
\frac{1}{U}=\frac{1}{2N} \sum_{\alpha, k>0}\left \{\frac{E_{\alpha k}^*+\left [\xi_k-(-1)^\alpha\Gamma_{BS}\right]}{\left(\left|u_{\alpha k}\right|^2+\left|v_{\alpha k}\right|^2\right)\left|E_{\alpha k}\right|\left|E_{\alpha k}+\left [\xi_k-(-1)^\alpha\Gamma_{BS}\right]\right|}\right \}.
\end{equation}
Similarly, the gap equations derived from $\bar{\Delta}$ and $\Delta$ are complex conjugates of each other and are therefore equivalent.

The presence of backscattering does not alter the definition of the order parameters. As a result, the relation described in Eq.~\eqref{equ:S1:eqorderphase} remains valid, thereby ensuring that $\Delta_0$, $E_k$, and the gap equation preserve gauge invariance.
In the weak scattering limit ($\Gamma_{BS} \to 0$), Eq.~\eqref{equ:S1:eqBSgapeq} reduces to Eq.~\eqref{equ:S1:eqgapeq}. Furthermore, in the Hermitian limit ($\gamma =0$), we have $\Delta_0 \in \mathbb{R}$ and $E_{\alpha k}=\sqrt{\left[\xi_k-(-1)^\alpha \Gamma_{BS}\right]^2+\Delta_0^2}$, simplifying Eq.~\eqref{equ:S1:eqBSgapeq} to the conventional Hermitian form
\begin{equation}
\frac{1}{U}= \frac{1}{2 N} \sum_{\alpha, k>0} \frac{1}{E_{\alpha k}}.
\end{equation}

\section{Comparison with the biorthogonal counterpart}
In this section, we derive the gap equation for the complex-valued interaction superfluid model within the biorthogonal approach \cite{Yamamoto2019} and conduct a comparative analysis with the right-eigenstate-based approach. 
In NH quantum mechanics, the biorthogonal basis associated with a given Hamiltonian $H$ is defined by
\begin{equation}
\label{equ:S1:eqbio}
\begin{aligned}
& H\left|\Psi_n\right\rangle_R=E_n\left|\Psi_n\right\rangle_R, \\
& \hat{H} ^{\dagger}\left|\Psi_m\right\rangle_L=E_m^*\left|\Psi_m\right\rangle_L,
\end{aligned}
\end{equation}
where $\left|\Psi_n\right\rangle_R$ represents the right eigenstates, and $\left|\Psi_m\right\rangle_L$ represents the left eigenstates \cite{Brody2013}. These states satisfy the biorthogonality condition ${ }_L\left\langle\Psi_m \mid \Psi_n\right\rangle_R=\delta_{m n}$. 

Within the biorthogonal-eigenstate-based NH mean-field theory, the order parameters for the complex-valued interaction superfluid model are defined as
\begin{equation}
\label{equ:S1:eqbioD}
\begin{aligned}
& \Delta^{LR}=-\frac{U}{N} \sum_k{ }_L\left\langle\Psi_0\right| c_{-k \downarrow} c_{k \uparrow}\left|\Psi_0\right\rangle_R, \\
& \bar{\Delta}^{LR}=-\frac{U}{N} \sum_k{ }_L\left\langle\Psi_0\right| c_{k \uparrow}^{\dagger} c_{-k \downarrow}^{\dagger}\left|\Psi_0\right\rangle_R,
\end{aligned}
\end{equation}
where $\left|\Psi_0\right\rangle_R$ and $\left|\Psi_0\right\rangle_L$ are the biorthogonal NH BCS ground states. 

Following the mean-field approximation and the NH Bogoliubov transformation as detailed in Section \ref{subsec:two_body_loss}, the biorthogonal NH BCS ground states are given by
\begin{equation}
\label{equ:S1:eqbioBSC}
\begin{aligned}
& \left|\Psi_0\right\rangle_R=\prod_k \gamma_{k \uparrow} \gamma_{-k \downarrow}|0\rangle=\prod_k\left(u_k^{LR}+v_k^{LR} c_{k \uparrow}^{\dagger} c_{-k \downarrow}^{\dagger}\right)|0\rangle, \\
& \left|\Psi_0\right\rangle_L=\prod_k \bar{\gamma}_{k \uparrow}^{\dagger} \bar{\gamma}_{-k \downarrow}^{\dagger}|0\rangle=\prod_k\left(u_k^{LR^*}+\bar{v}_k^{LR^*} c_{k \uparrow}^{\dagger} c_{-k \downarrow}^{\dagger}\right)|0\rangle,
\end{aligned}
\end{equation}
where 
\begin{equation}
u_k^{LR}=\sqrt{\frac{E_k^{LR}+\xi_k}{2 E_k^{LR}}}, \ \  v_k^{LR}=-\sqrt{\frac{E_k^{LR}-\xi_k}{2 E_k^{LR}}} \sqrt{\frac{\Delta^{LR}}{\bar{\Delta}^{LR}}},\ \  \bar{v}_k^{LR}=-\sqrt{\frac{E_k^{LR}-\xi_k}{2 E_k^{LR}}} \sqrt{\frac{\bar{\Delta}^{LR}}{\Delta^{LR}}},
\end{equation}
and $\bar \gamma_{ k\sigma}$, $\gamma_{ k\sigma}$ are the creation and annihilation operator of quasiparticles, $E_{k}^{LR}= \sqrt {\xi_k^2 + \Delta^{LR} \bar{\Delta}^{LR}}$ are the quasiparticle excitation energies.

By substituting Eq.~\eqref{equ:S1:eqbioBSC} into Eq.~\eqref{equ:S1:eqbioD}, we derive the NH gap equation
\begin{equation}
\label{equ:S1:eqbiogapequ}
\frac{1}{U}= \frac{1}{2 N} \sum_k \frac{1}{E_k^{LR}}.
\end{equation}
This equation exhibits a concise mathematical form, analogous to the Hermitian case.
Considering the U(1) symmetry and the mean-field treatment in Eq.~\eqref{equ:S1:eqbioD}, we can generally define
\begin{equation}
\label{equ:S1:eqorderbio}
\begin{aligned}
\Delta^{LR}=\Delta_0^{LR} e^{i \phi}, \ \ \ \bar{\Delta}^{LR}=\Delta_0^{LR} e^{-i \phi},
\end{aligned}
\end{equation}
where $\Delta_0^{LR} \in \mathbb{C}$. 
Under this definition, $\Delta^{LR} \bar{\Delta}^{LR}=(\Delta_0^{LR})^2$ remains gauge-invariant. Therefore, $E_k^{LR} $ and the gap equation [Eq.~\eqref{equ:S1:eqbiogapequ}] are also independent of the gauge phase $\phi$.

The numerical solutions of Eq.~\eqref{equ:S1:eqgapeq} and Eq.~\eqref{equ:S1:eqbiogapequ}, computed under identical parameters, are presented in Fig.~\ref{fig:figs}. In the Hermitian limit, these solutions converge. However, the introduction of two-body loss leads to a significant divergence in their behaviors. Specifically, the biorthogonal approach (LR) fails to produce nontrivial solutions within the regime of moderate dissipation and exhibits discontinuities. In contrast, the right-eigenstate-based approach (RR) generates continuous solutions and achieves a lower condensation energy, underscoring its superior suitability for accurately describing the system under investigation.
Indeed, from the perspective of experimental observability, a fundamental distinction arises between these two definitions. The right-eigenstate-based definition not only provides a coherent probabilistic interpretation but also aligns more closely with experimental measurements, thereby enhancing its practical relevance \cite{Meden2023,Graefe2008,Dora2019,Sticlet2022,Yamamoto2023,Yan2024,Kornich2023,Yamamoto2022,Barontini2013,Li2019,Wu2019}. Our investigation further illustrates this by providing a concrete example within open superfluid systems.

%figure in one column
\begin{figure*}[hbt] 
\centering
\includegraphics[width=0.68\textwidth]{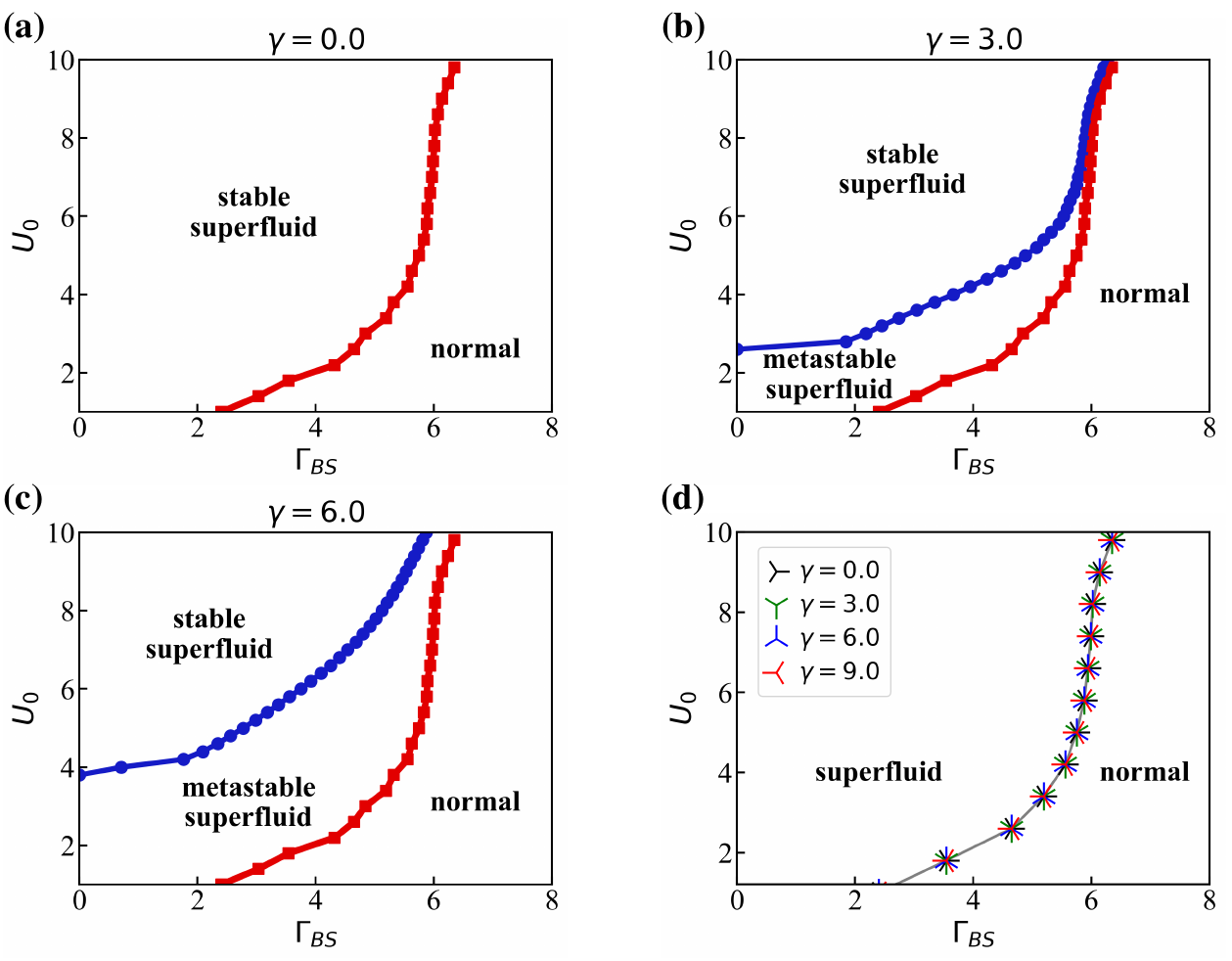} 
\caption{Phase diagrams in the $U_0 - \Gamma_{BS}$ plane: (a) $\gamma=0$; (b) $\gamma=3$; (c) $\gamma=6$. These diagrams correspond to cross-sections along the $\gamma$ axis in Fig.~\ref{fig:fig4}(b). The blue curve denotes the condition $\text{Re}E_c = 0$ with $\Delta_0 \ne 0$, while the red curve marks the onset of $\Delta_0 = 0$. (d)  The critical lines representing the transition to the normal phase (corresponding to the red curves in panels (a), (b), and (c)) exhibit complete overlap across various values of $\gamma$.}
\label{fig:figs1}
\end{figure*}

\section{Phase diagram in the presence of backscattering}
By analyzing the order parameter and condensation energy, we construct the phase diagram in the $U_0$-$\Gamma_{BS}$ plane at different $\gamma$, as shown in Fig.~\ref{fig:figs1}(a), (b), and (c).
In the Hermitian limit ($\gamma=0$), sufficiently strong backscattering completely destroys superfluidity, driving a superfluid-to-normal phase transition, as illustrated in Fig.~\ref{fig:figs1}(a).
This disruptive effect persists in the NH superfluid system with two-body loss and remains independent of the loss parameter $\gamma$.
As shown in Fig.~\ref{fig:figs1} (d), the critical lines of the transition to the normal phase remain invariant across different values of $\gamma$.
Moreover, two-body loss destabilizes the superfluid state, rendering it metastable, while backscattering further amplifies this destabilization. As depicted in Fig.~\ref{fig:figs1}(b) and (c), the parameter space corresponding to the stable superfluid phase shrinks with increasing $\gamma$ and $\Gamma_{BS}$.

In summary, the phase diagrams reveal the intricate interplay between pairing interaction, two-body loss, and backscattering. While the pairing interaction serves to stabilize the superfluid phase, two-body loss and backscattering act as destabilizing factors. The competition among these effects gives rise to a distinct metastable superfluid phase in the NH superfluid systems.

\end{document}